\documentclass[%
reprint,
superscriptaddress,
frontmatterverbose,
amssymb,amsfonts,amsmath,
footinbib,
pre,
nofootinbib
]{revtex4-1}

\usepackage{times}
\usepackage{graphicx}
\usepackage{dcolumn}
\usepackage{bm}
\usepackage{epsfig}
\usepackage{epstopdf}
\usepackage{natbib}
\usepackage[version=4]{mhchem}
\usepackage{soul}
\usepackage[mathlines]{lineno}
\usepackage{color,soul}
\usepackage[hidelinks=true]{hyperref}
\hypersetup{colorlinks   = true, urlcolor     = blue, linkcolor    = blue,  citecolor   = red }

\newcommand{\bra}[1]{\left(#1\right)}

\newcommand{\rev}{\color{red}} 

\begin{document}
	
	\title{Molding the asymmetry of localized frequency-locking waves by a generalized forcing and implications to the inner ear}
	
	\author{Yuval Edri}
	\affiliation{Department of Solar Energy and Environmental Physics, Swiss Institute for Dryland Environmental and Energy Research, Blaustein Institutes for Desert Research (BIDR), Ben-Gurion University of the Negev, Sede Boqer Campus, 8499000 Midreshet Ben-Gurion, Israel}%
	\author{Dolores Bozovic}
	\affiliation{Department of Physics and Astronomy and California NanoSystems Institute, University of California Los Angeles, Los Angeles, CA, 90025, USA}%
	\author{Ehud Meron}%
	\affiliation{Department of Solar Energy and Environmental Physics, Swiss Institute for Dryland Environmental and Energy Research, Blaustein Institutes for Desert Research (BIDR), Ben-Gurion University of the Negev, Sede Boqer Campus, 8499000 Midreshet Ben-Gurion, Israel}%
	\affiliation{Department of Physics, Ben-Gurion University of the Negev, 8410501 Beer Sheva, Israel}%
	\author{Arik Yochelis}%
	\email{yochelis@bgu.ac.il}
	\affiliation{Department of Solar Energy and Environmental Physics, Swiss Institute for Dryland Environmental and Energy Research, Blaustein Institutes for Desert Research (BIDR), Ben-Gurion University of the Negev, Sede Boqer Campus, 8499000 Midreshet Ben-Gurion, Israel}%
	\affiliation{Department of Physics, Ben-Gurion University of the Negev, 8410501 Beer Sheva, Israel}%

	\date{\today}
	
	\begin{abstract}
		
	Frequency locking to an external forcing frequency is a {well} known phenomenon. In the auditory system, it results in a localized traveling wave, the shape of which is essential for efficient discrimination between incoming frequencies. An amplitude equation approach is used to show that the shape of the localized traveling wave depends crucially on the relative strength of additive vs. parametric forcing components; the stronger the parametric forcing the more asymmetric the response profile and the sharper the traveling-wave front. The analysis captures the empirically observed regions of linear and nonlinear responses and highlights the significance of parametric forcing mechanisms in shaping the resonant response in the inner ear.
		
	\end{abstract}
	
	\maketitle
	
	
Frequency-locking phenomena have been observed in a wide variety of oscillatory systems that are subjected to temporal periodic forcing, including liquid crystals~\cite{coullet1994excitable}, nonlinear optical systems~\cite{tlidi1998kinetics,izus2000bloch}, photosensitive chemical reactions~\cite{marts2006resonant}, and autocatalytic surfaces~\cite{imbihl1995oscillatory}. Despite the essential differences in their physicochemical properties, these systems share similar spatio-temporal behaviors, as, close to the onset of oscillations, they can all be reduced to the same normal form -- the forced complex Ginzburg-Landau (FCGL) equation~\cite{CoulletEmilsson1992} and variants thereof~\cite{rudiger2007theory}. The simplest of these behaviors are resonant uniform oscillations in which the actual oscillation frequency is locked to a simple fraction of the forcing frequency, in a range of the latter that depends on the forcing amplitude~\cite{marts2006resonant}. More intricate behaviors include resonant standing and traveling-wave patterns~\cite{Lega1990,nib1997,elphick1998phase,yochelis2002development}, and localized resonant oscillations~\cite{BurkeYochelisKnobloch2008,ma2012depinning,tzou2013homoclinic,castiloyochelis}.
	
Frequency-locking is fundamental to the auditory system, as the sense of hearing in the inner ear achieves extremely high sensitivity of detection, as well as frequency discrimination, by entrainment to incoming sound waves. An incoming sound elicits a localized traveling wave (TW) along the basilar membrane~\cite{zweig1976basilar,lighthill1981energy,FisherNinReichenbachEtAl2012} {that extends throughout the snail-shaped cochlea. In mammals the localization is largely attributed to the spatial inhomogeneity of the cochlea and the associated monotonic gradient of natural frequencies, as high-amplitude resonant responses occur only at locations where the incoming frequencies match the local natural frequencies~\cite{von_bekesy,ReichenbachHudspeth2014,warren2016minimal}. The TW localization constitutes the basis for sound discrimination, as different incoming frequencies trigger localized TW at different locations along the cochlea. In the mammalian cochlea, high frequencies are detected at the organ’s base, while low frequencies maximally stimulate the apical locations. Furthermore, the localized TW exhibits an asymmetric waveform at the basal locations, and a more symmetric envelope at the apex~\cite{siebert1968stimulus,zweig1976basilar}. The interaction between the mechanical membrane oscillations and local sensory cells---the so-called hair cells~\cite{gold1948hearing,AshmoreKolston1994,MartinHudspeth1999,CamaletDukeJulicherEtAl2000,RoblesRuggero2001,essential,neyman2010,gelfand2010,reichenbach2010ratchet,fredrickson2012,levy2016high}, results in the release of neurotransmitters and in neuronal signaling~\cite{dallos1996overview}.}
	
Several theoretical studies have modeled the auditory system using the normal form equation for {forced oscillations, i.e. the the FCGL equation}~\cite{SzalaiChampneysHomerEtAl2013,Hudspeth2014} and the references therein. The rational behind this approach, which does not model specific mechanical and electrophysiological processes, is the focus on universal aspects of forced inhomogeneous oscillatory systems, which nevertheless can be related to specific processes once a faithful model for the complex auditory system becomes available. The models were shown to capture observed responses of the auditory hair cells to incoming sound waves~\cite{essential,ourepl}, including amplification, frequency selectivity, and the spatial localization of the TW response~\cite{kern2003,Magnasco2003,duke_julicher2003,ReichenbachHudspeth2014}. Despite this success, an important aspect of the interaction between the incoming sound wave and the cochlear response has been overlooked -- the inherent asymmetry in the spatial profile of the localized TW, suggested to be relevant to the discrimination between incoming frequencies ~\cite{temchin2008threshold,temchin2014spatial,warren2016minimal}. Although {these studies have shown} reasonable fits to empirical data~\cite{zweig1976cochlear,steele1979comparison,babivc1983vibration,leveque1988solution,kern2003,Magnasco2003,duke_julicher2003,grosh2010,reichenbach2010ratchet,stockei}, the factors that affect this asymmetry {have remained elusive}.
	
In this Letter, we show that the asymmetry of the frequency-locked response is strongly affected by the form of the driving periodic force. The coupling of an external driving force to state variables of an oscillatory system can be additive or parametric. The former coupling form is typical to mechanical systems~\cite{Nettel2009}, whereas the latter is often found in {reaction-diffusion-type} systems~\cite{Lin2004pre}. The complex auditory system contains both mechanical elements, associated with the basilar membrane and with hair-cell bundles, and {reaction-diffusion} elements (a.k.a. activator-inhibitor), associated with the openings of ion channels in hair cells~\cite{Ashmore1985,neyman2010,ourepl}. A deep understanding of the response of the auditory system to incoming sound waves therefore calls for the consideration of both additive forcing~\cite{SzalaiChampneysHomerEtAl2013,ReichenbachHudspeth2014,Hudspeth2014} and the hitherto overlooked parametric forcing.
	
We address the combined effect of the two forcing forms using a generalized FCGL equation for an over-damped inhomogeneous oscillatory medium that is subjected to a TW forcing, mimicking phenomenologically the frequency-locking dynamics in the cochlea. We focus on the interplay between additive and parametric TW forcing and how it affects the spatial profile of the localized wave response. Our results not only highlight the importance of the parametric component of the forcing in inducing strong asymmetry, as observed in experiments, but also demonstrate the essential role it plays in capturing the different linear and nonlinear response regimes that have been observed at increasing intensity levels of incoming sound waves~\cite{MartinH01,ruggero97}.
	
{Viewing the cochlea as a spatially extended oscillatory system near the onset of a spatially-uniform Hopf bifurcation~\cite{hudspeth2010}, observables of that system can be expressed as
	\begin{equation*}
		{\mathbf{u}}(x,t) \propto  Be^{i\omega_c t}+\text{complex conjugate}  \,,
		\end{equation*}
		where $\omega_c$ is the {natural frequency of the oscillations that appear} at the bifurcation point (Hopf frequency), and $B$ is a complex-valued amplitude, assumed to be small in absolute value and slowly varying in space and time, that satisfies the well known complex Ginzburg-Landau (CGL) equation~\cite{aranson2002world}. The observables $\mathbf{u}(x,t)$ represent physiological variables, such as basilar-membrane deformations, fluid pressure, and neural activity. Resonant 1:1 entrainment of the system by the incoming TW, evoked by combined additive and parametric forcing, is described by {constant solutions of} the generalized FCGL equation } ~\cite{rudiger2007theory,ourepl}:
	\begin{eqnarray}\label{eq:FCGL}
	\nonumber	\frac{\partial A}{\partial t}&=&\left(\mu+i\nu\right)
	A-\left(1+i\beta\right)|A|^2A+\gamma\bra{\Gamma_a+\gamma \Gamma_{p} \bar A} \\
	&&+\left(1+i\alpha\right)\left(\frac{\partial^2 A}{\partial x^2}-ik_f\frac{\partial A}{\partial x}\right)\,,
	\end{eqnarray}
{where $A$ is the amplitude of a traveling-wave form and $\bar A$ is complex conjugate (see Supplemental Material at \cite{supplementalMaterial} for details). In Eq.~\eqref{eq:FCGL} $\mu$ is the distance from the Hopf bifurcation, $k_f$ is the wavenumber of the TW forcing, $\nu=\omega_c-\omega_f$ represents the deviation of the forcing frequency, $\omega_f$, from exact resonance, or the detuning, and $\alpha$ and $\beta$ are real parameters related to dispersive and nonlinear corrections of the oscillation frequency, respectively. The parameter $\gamma$ accounts for the correlated forcing strength of both the additive and parametric forcing, while $\Gamma_a$ and $\Gamma_p$ control the relative strength of the respective forcing type (see Supplemental Material at \cite{supplementalMaterial} for details).
Note that the advective term $\partial_x A$ in~\eqref{eq:FCGL} cannot be eliminated via a co-moving frame transformation, since the coefficient of this term is complex-valued and thus acts as differential advection~\cite{chomaz1992absolute}.
	
{Motivated by empirical observations~\cite{von_bekesy}, we introduce the spatial inhomogeneity of the cochlea through an exponentially decreasing critical Hopf frequency, $\omega_c(x)=\omega_{0}\exp(-\kappa x)$, where $\kappa = L^{-1}\ln(\omega_0/\omega_L)$ and $\omega_{0}$ and $\omega_{L}$ are, respectively, the frequencies at the cochlea base, $x=0$, and at the cochlea apex, $x=L$.}~\cite{duke_julicher2003}.
	\begin{figure}[t]
		(a)\includegraphics[width=0.4\textwidth]{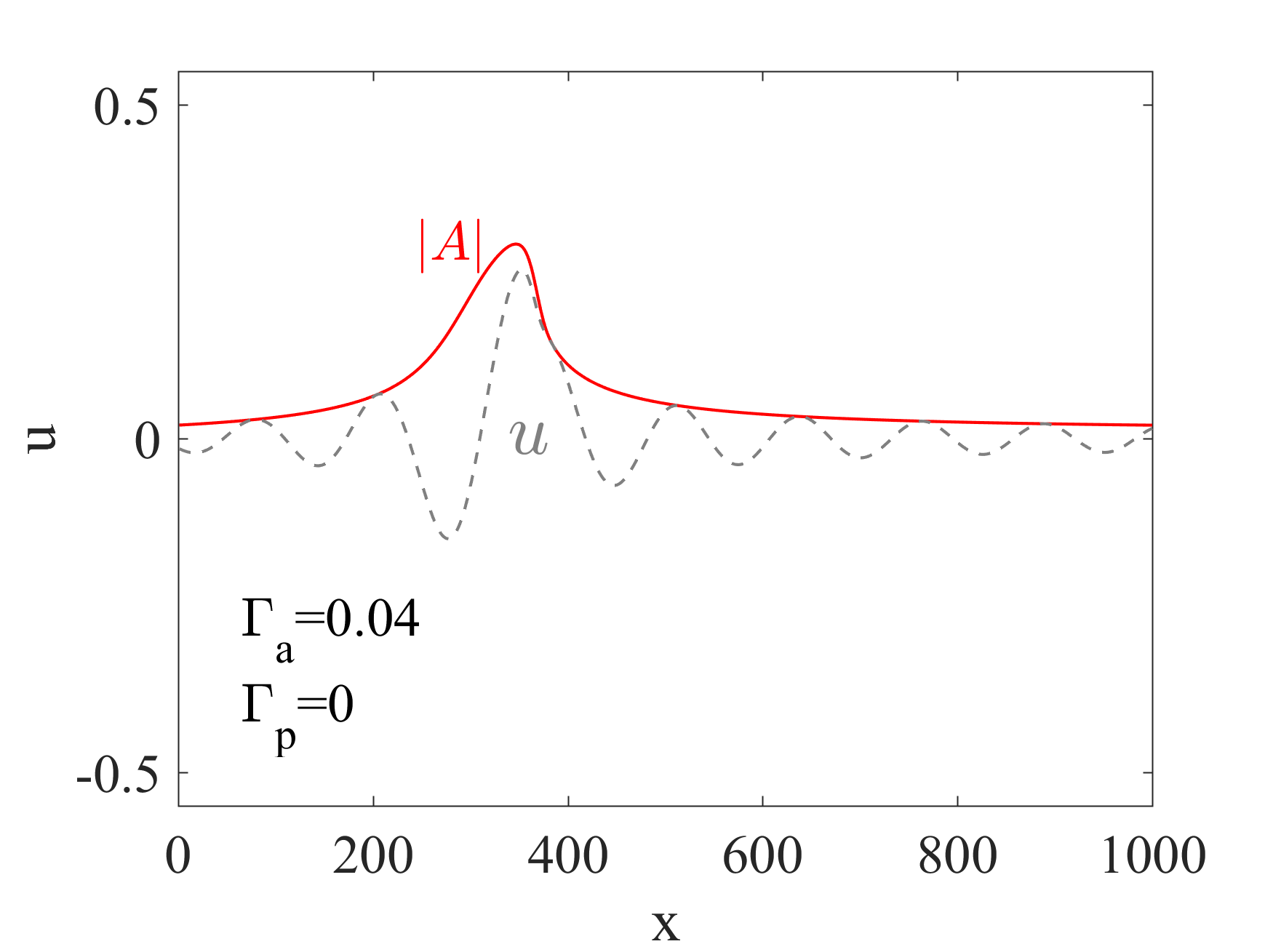}
		(b)\includegraphics[width=0.4\textwidth]{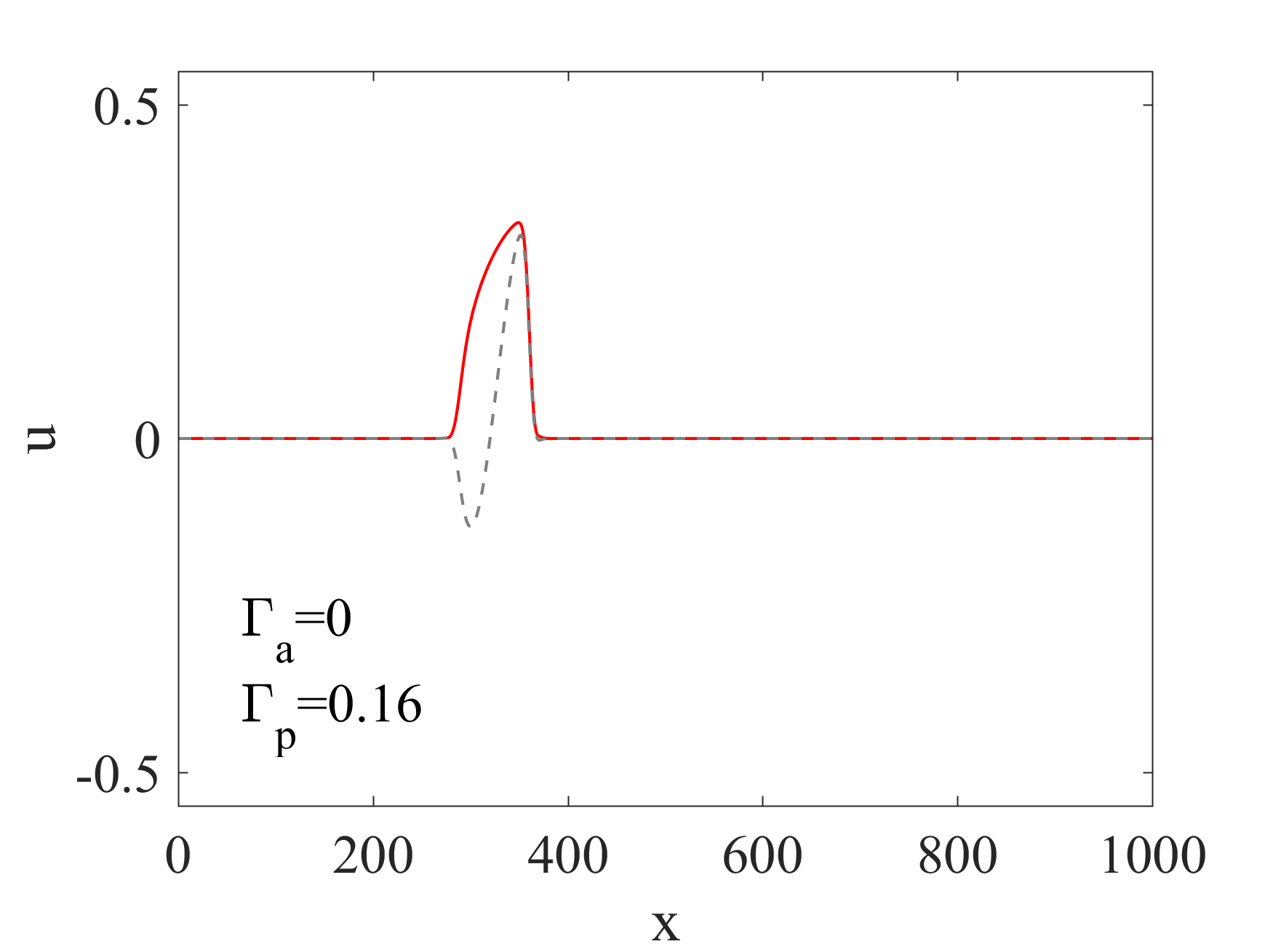}
		(c)\includegraphics[width=0.4\textwidth]{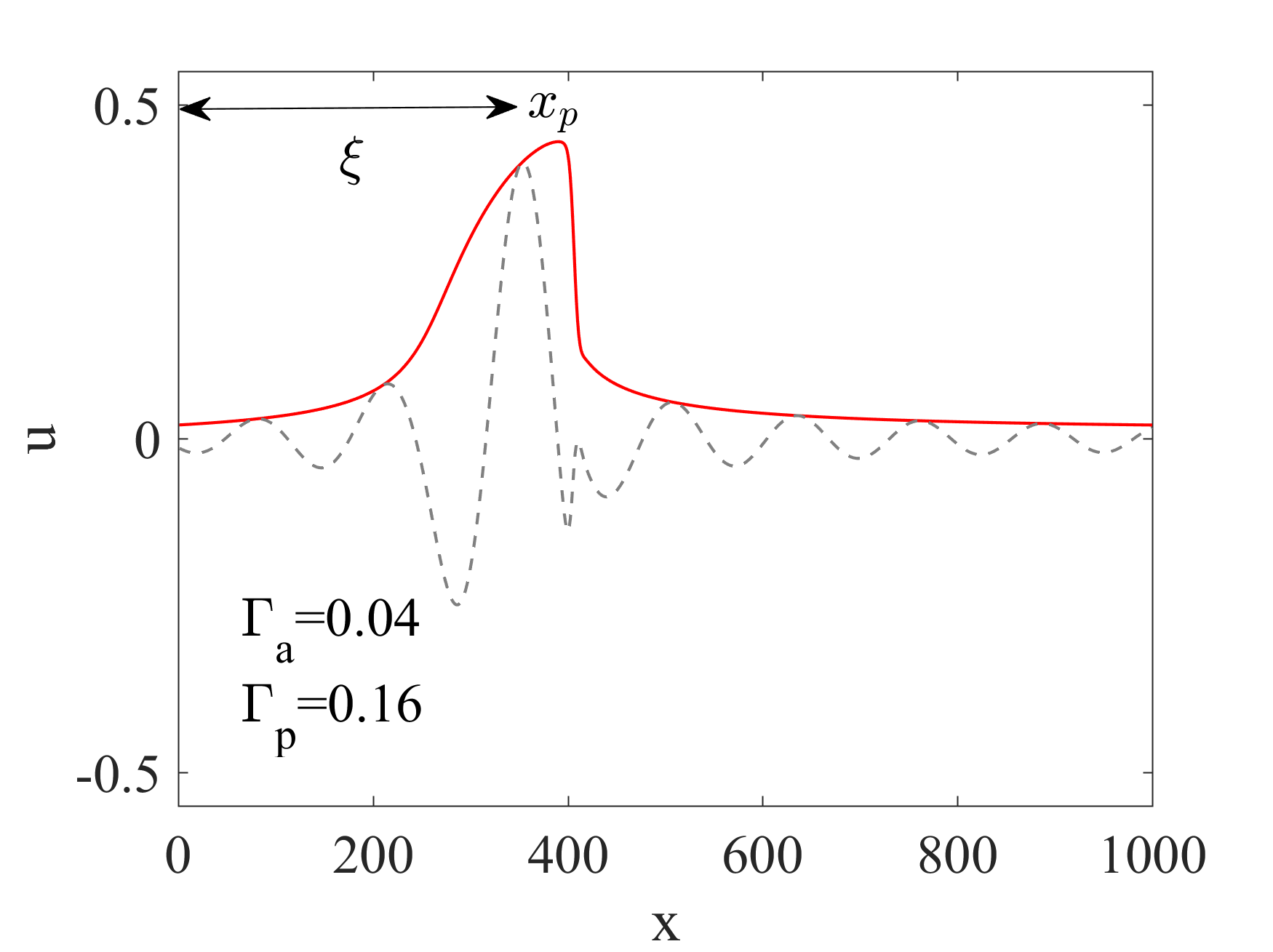}
		\caption{Localized traveling-wave solutions of the form $u\propto |A|\cos{(\omega_ft-k_fx)}$ (dashed lines) using Neumann boundary conditions and their envelopes (solid lines), at different {proportions} of (a) additive, (b) parametric, and (c) combined forcing components. {In (c) $x_p$ marks the location of {the TW} peak and $\xi$ is the minimal distance to the boundaries (see Eq. \eqref{fig:Lambda} and the text that follows it for details).} {Parameters: $\mu=-0.0525$, $\gamma=1$, $\beta=-1.7$, $\alpha=0$, $k_f=16\pi/L$, $\omega_{0}=5$, $\omega_{L}=1$, $\omega_f=3$, $L=1000$.}}
		\label{fig:yochelis4beta17v2}
	\end{figure}	
	\begin{figure}[t]
		\centering
		\includegraphics[width=0.45\textwidth]{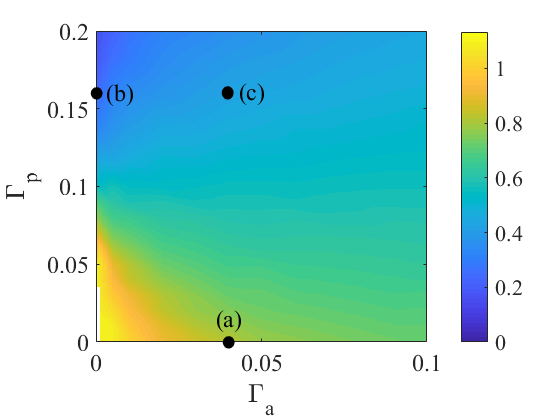}
		\caption{Asymmetry map for localized frequency-locked solutions. Shown is the asymmetry measure $\Lambda$, defined by Eq. ~\eqref{eq:sym}, as a function of the additive and parametric forcing components, $\Gamma_a$ and  $\Gamma_p$, respectively, where $\Lambda \to 0$ corresponds to maximal asymmetry. Both forcing forms increase asymmetry but the effect of parametric forcing is stronger. The black dots corresponds to the parameters used in Fig.~\ref{fig:yochelis4beta17v2}, respectively. }
		\label{fig:Lambda}
	\end{figure}
{To study the impact} of combined additive and parametric forcing on localized frequency locked solutions, we first solved~\eqref{eq:FCGL} using direct numerical integration and obtained the localized response forms shown in Fig.~\ref{fig:yochelis4beta17v2}(a-c). {For both purely additive forcing ($\Gamma_p=0$) and purely parametric forcing ($\Gamma_a=0$), maximal response is achieved at a specific location, where the forcing frequency $\omega_f$ matches the space-dependent natural frequency $\omega_c(x)$, as Fig.~\ref{fig:yochelis4beta17v2}(a,b) shows. However, there is an essential difference between the responses to the two forcing types; while for  purely additive forcing the response extends over the entire domain, even though the amplitude is low, for purely parametric forcing the response is limited to a narrow region and vanishes otherwise (see Ref. ~\cite{ourepl} for a detailed discussion of resonant responses in uniform systems). Another aspect that distinguishes between the two forcing types is the spatial profile of the response, being highly asymmetric in the parametric case, and fairly symmetric in the additive case.} The combined forcing, including both terms, results in an increase of both profile asymmetry and spatial width with respect to the pure parametric case, as shown in Fig.~\ref{fig:yochelis4beta17v2}(c).
	
To quantify the profile asymmetry, we define a measure that depends on the amplitude's absolute value $\rho=|A|$:
	\begin{equation}\label{eq:sym}
	\Lambda =\int_{x_p }^{x_p+\xi} {\rho {\kern 1pt} {\rm{d}}x} {\bigg /} \int_{x_p-\xi}^{x_p} {\rho {\kern 1pt} {\rm{d}}x} ,
	\end{equation}
where $x_p$ is the location of the response peak, i.e., $\rho_{|x=x_p}= \max (\rho) \equiv \rho_{\max}$, and $\xi$ is the minimal distance from $x_p$ to one of the boundaries, $\xi\equiv \min(x_p,L-x_p) \sim \mathcal{O}(L/2)$, where $L$ is the length of the system. Hence, the profile properties range from $\Lambda \to 0$ for strongly asymmetric localization to $\Lambda \to 1$ for symmetric localization. The asymmetry of the response, in terms of this measure, for various combinations of additive and parametric forcing components is shown in Fig.~\ref{fig:Lambda}.
	
The strong asymmetry induced by the parametric-forcing component can be understood by relating the space-dependent Hopf frequency, $\omega_c(x)$, to the forcing frequency, $\omega_f$, {along the} $x$ axis. {Studies of frequency locking in parametrically-forced overdamped oscillatory systems~\cite{lifshitz,ourepl} indicate that the bifurcation from a stationary state to resonant oscillations, as $\omega_c=\omega_f$ is approached, is subcritical when the detuning $\nu=\omega_c-\omega_f$ is negative (for $\beta<0$), implying an abrupt transition to oscillations, and supercritical when the detuning is positive (see Fig. ~\ref{fig:bif}(a,b))}. Since the detuning decreases from positive to negative values as a resonance is traversed along the $x$ axis (because of the spatial dependence of the Hopf frequency), the {amplitude of the localized TW increases smoothly but decreases abruptly, as the superimposed solution in Fig.~\ref{fig:bif}(b) shows. Thus, the strong asymmetry of the profile in the purely parametric-forcing case is a consequence of the subcritical nature of the bifurcation at negative detuning. The absence of long tails on both sides of the localized TW is a consequence of the stability of the stationary state outside the domain of resonant oscillations and the exponential decay of the oscillation amplitude to zero. The position and width of the TW localization is determined by the limited frequency-locking range.}
	
	\begin{figure*}[tp]
		\centering
		{
			(a)\includegraphics[width=0.45\textwidth]{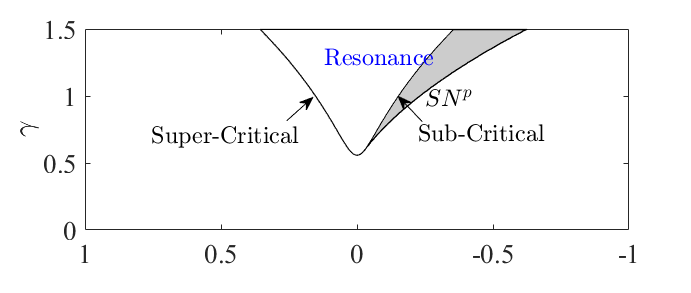}
			(c)\includegraphics[width=0.45\textwidth]{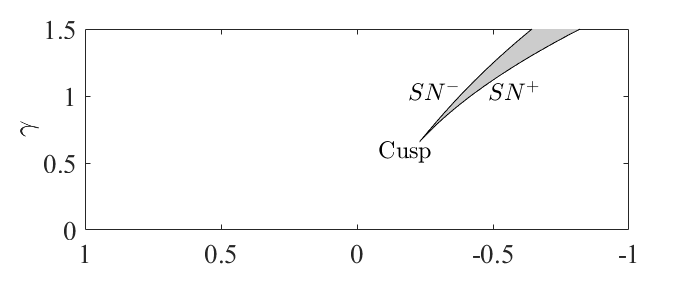}
			(b)\includegraphics[width=0.45\textwidth]{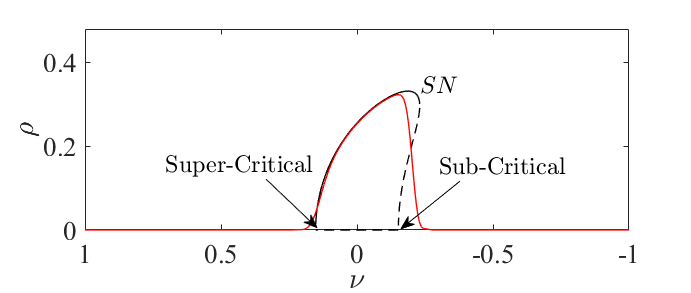}
			(d)\includegraphics[width=0.45\textwidth]{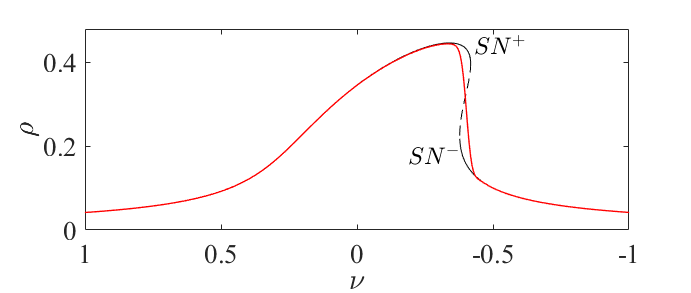}
		}
		\caption{{Resonant response to  {parametric forcing with $(\Gamma_a,\Gamma_p)=(0,0.16)$ (panels a,b), and to combined forcing $(\Gamma_a,\Gamma_p)=(0.04,0.16)$ (panels c,d). In the parametric-forcing case, resonant oscillations are limited to a tongue-like domain in the $(\nu,\gamma)$ plane as panel (a) shows. The shaded area denotes bistability of the trivial (zero-amplitude) state and of resonant oscillations. It is bounded by the sub-critical bifurcation of the trivial state and by the saddle-node bifurcation ($SN^p$) of resonant oscillations, as the bifurcation diagram (black line) in panel (b) shows for $\gamma=1$. Panel (b) also shows a superimposed envelope {\rev $|A|$} of the localized TW solution (see Fig.~\ref{fig:yochelis4beta17v2}(b)), expressed as a function of $\nu(x)$ (red line). In the combined-forcing case, resonant solutions persist for any detuning $\nu$, as the bifurcation diagram (black line) in panel (d) shows for $\gamma=1$. The shaded area in panel (c) denotes bistability of high-amplitude and low-amplitude oscillatory solutions, appearing in a pair of saddle-node bifurcations, $SN^+$ and $SN^-$, respectively. Panel (d) also shows a superimposed  localized TW solution (red line).} Other parameters are as in Fig.~\ref{fig:yochelis4beta17v2}.}}
		\label{fig:bif}
	\end{figure*}
	
The main effect of an additive-forcing component ($\Gamma_a>0$) is to destroy the stationary, zero-amplitude state and replace it by resonant oscillations of low amplitude, which increases in size as the resonance is approached. As a result, the localized TW acquires long tails that extend to the entire system, unlike the strictly localized TW in the case of pure parametric forcing (compare panels (b) and (c) in Fig.~\ref{fig:yochelis4beta17v2}). While the combined additive-parametric forcing breaks the strong locality of the TW profile in the case of pure parametric forcing, it does keep the asymmetry of the profile. In fact, even in the absence of a parametric component ($\Gamma_p=0$) the TW profile may become asymmetric. This asymmetry is associated with a cusp singularity that develops for positive detuning, as shown by the $(\nu,\gamma)$ parameter space in Fig.~\ref{fig:bif}(b). The singularity involves the appearance of a pair of saddle-node bifurcations which create bistability of low-amplitude and high-amplitude resonant oscillations and an abrupt decline of the oscillation amplitude as the detuning increases along the $x$ axis, as Fig.~\ref{fig:bif}(b,d) shows. For a pure additive forcing, the cusp singularity exists for $\beta>\sqrt{3}$~\cite{yipingtheis}, while for combined forcing it may exist for any $|\beta|>0$. We note that while frequency-locking solutions for either purely parametric or purely additive forcing can be obtained analytically~\cite{ourepl}, the case of combined forcing requires a numerical approach. The results, shown in Fig.~\ref{fig:bif}, were obtained by numerical continuation, using the package AUTO~\cite{auto} and standard linear stability analysis.
	
	\begin{figure}[t]
		\centering
		\includegraphics[width=1\linewidth]{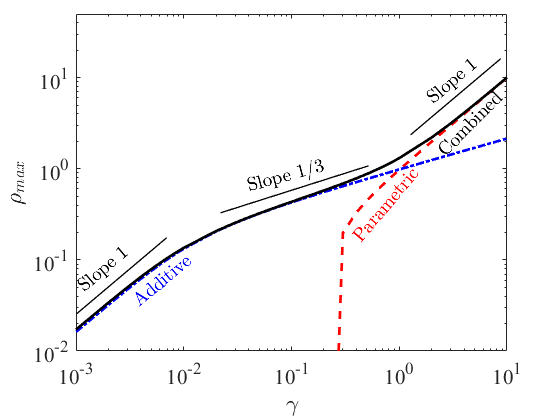}
		\caption{Nonlinear response of the oscillation amplitude to combined forcing. Shown is a log-log plot of the maximal oscillation amplitude, obtained by solving Eq.~\eqref{eq:FCGL}, as a function of combined-forcing strength $\gamma$ {and with $\Gamma_a=\Gamma_p=1$}. For comparison, the responses to purely additive forcing  ({$\Gamma_a=1,\Gamma_p=0$}, dash-dotted line) and to purely parametric forcing ({$\Gamma_a=0,\Gamma_p=1$}, dashed line) are also shown. {Other parameters: $\beta=0$ and the rest are as in Fig.~\ref{fig:yochelis4beta17v2}.}}
		\label{fig:compression}
	\end{figure}
	
The observed response of the basilar membrane and of individual hair cells to increasing stimulus intensities can be divided into three regimes of scaling relations between the oscillation amplitude and the stimulus~\cite{ruggero97,Martin01,MartinH01}: a linear relation at low intensities, a nonlinear, cubic-root relation at moderate intensities (compressed response), and a linear relation again at high intensities. Combined forcing readily reproduces all three regimes, as Fig.~\ref{fig:compression} shows, while either a purely additive forcing (dashed-dot line) or a purely parametric forcing (dashed line) fail to do so, as reviewed in more detail in~\cite{SzalaiChampneysHomerEtAl2013}.
	
In this study, we disentangled the response of an inhomogeneous overdamped oscillatory system to traveling-wave (TW) forcing, by phenomenologically studying the effects of parametric vs. additive forcing. We focused on the auditory system as a particularly interesting application, where the actual shape of the localized TW response was found to be different along the cochlear domain~\cite{RoblesRuggero2001}. Using an amplitude equation approach we showed that purely parametric forcing results in an oscillation amplitude that sharply declines to zero away from the localized TW (Fig.~\ref{fig:yochelis4beta17v2}(b)), while additive, or combined additive and parametric response, results in long tails of low-amplitude oscillations (Fig.~\ref{fig:yochelis4beta17v2}(a,c)), which delocalize the TW response and thereby may affect the quality of sound discrimination. Furthermore, the spatial profile of the localized response in the case of pure parametric forcing is highly asymmetric -- a consequence of bistability of finite-amplitude and \emph{zero-amplitude} solutions in the leading edge of the TW response. Combined parametric and additive forcing may also result in a highly asymmetric response, except that this response is a consequence of bistability of a high-amplitude and low-amplitude (rather than zero-amplitude) solutions, and therefore is still accompanied by long tails.
	
{These results may shed new light on the mechanisms by which the auditory system responds to incoming sound waves along the cochlea.
Empirical observations indicate that the apical part of the cochlea exhibits broader and more symmetric TW profiles~\cite{temchin2008threshold}, as compared to those observed in the basal part. These observations can be accounted for, in our approach, by assuming a change in the relative strength of the additive and parametric components of the forcing along the cochlea, that is, by assuming that the ratio $\Gamma_p/\Gamma_a$ decreases along the cochlia axis $x$. This, in turn, suggests that feedback processes that can be associated with parametric forcing, possibly, ion transport in hair cells~\cite{neyman2010} and the decoupling of hair cells from the tectorial membrane~\cite{reichenbach2010ratchet}, are dominant at the basal part of the cochlea, while processes associated with additive forcing, e.g. mechanical oscillations evoked by the incoming sound, are dominant at the apical part.}
	
Since the results reported here are based on a universal amplitude equation approach we should expect them to be applicable to a variety of other contexts that share the basic elements of overdamped oscillations, periodic forcing and spatial inhomogeneity. An interesting example is plant communities in water-limited systems. Models of dryland vegetation predict the existence of damped oscillatory modes that describe oscillatory convergence to uniform vegetation~\cite{Klausmeier1999science,Tzuk2018ptrsa}. A plant community with trait-dependent natural frequencies subjected to seasonal rainfall periodicity, falls in the class of amplitude equations considered here, with physical space being replaced by trait space. A localized biomass distribution in trait space defines the community structure~\cite{Nathan2016j_ecology, Meron2016mb}, and the tails of this distribution describe rare species. The existence of long tails can be of crucial importance to the resilience of plant communities in variable environments. Unraveling the factors that affect these tails, using a similar analysis to that reported here, can be highly significant.
	
	We thank James Hudspeth (Rockefeller University) and Elizabeth Olson (Columbia University) for useful discussions, and acknowledge the financial support by the Adelis Foundation and the Israel Science Foundation under Grant No. 305/13.
	\providecommand{\noopsort}[1]{}\providecommand{\singleletter}[1]{#1}%

\end{document}